# Simulation Studies of A Phenomenological Model for the Assembly of Elongated Virus Capsids

Ting Chen[1] and Sharon C. Glotzer[1,2,*]

[1]Department of Chemical Engineering and [2]Department of Materials Science & Engineering

University of Michigan, Ann Arbor, Michigan 48109-2136

[*]To whom correspondence should be addressed. Email: sglotzer@umich.edu

**ABSTRACT** We extend our previously developed general approach (1) to study a phenomenological model in which the simulated packing of hard, attractive spheres on a prolate spheroid surface with convexity constraints produces structures identical to those of prolate virus capsid structures. Our simulation approach combines the traditional Monte Carlo method with the method of random sampling on an ellipsoidal surface and a convex hull searching algorithm. Using this approach we study the assembly and structural origin of non-icosahedral, elongated virus capsids, such as two aberrant flock house virus (FHV) particles and the prolate prohead of bacteriophage φ29, and discuss the implication of our simulation results in the context of recent experimental findings.

## INTRODUCTION

Viruses are entities that replicate themselves using only a few types of proteins in the host cell, yet possess a sophisticated defense mechanism against outside attacks from, e.g., the host cell's immune system and antiviral drugs. The virus capsid, which assembles from coat protein subunits and provides the necessary protection for its enclosed genetic material from the environment, is a crucial functional component of any virus. Virus capsids can take various shapes. About half of all virus capsids are spherical and most of those have their protein capsomers arranged with icosahedral symmetry. Self-assembly of icosahedral viruses and the origin of their precise packings has received substantial attention in recent years (1-7). A good review on the theoretical progress of icosahedral viruses can be found in reference (8).

At the same time, the efficient packing of disks and spheres in 2D and 3D space, respectively, is a historical problem that continues to receive attention. In 3D space, it was recently shown that ellipsoids could pack more efficiently and could reach higher random packing fraction or jammed disordered packing fraction than regular spheres (9), showing the nontrivial influence of object shape on packing. In a 2D plane, the densest packing for spheres is known to be hexagonal close packing (HCP). Sphere packings on a spherical surface, a quasi-2D surface, have also been studied extensively (10) and were recently related to precise packings in icosahedral viruses (1, 4, 5). While significant progress has been achieved in the study of packings on spherical surfaces, the corresponding problem on non-spherical surfaces, e.g. ellipsoidal surfaces, has received little attention.

Furthermore, compared with the progress that has been made for understanding the origin of icosahedral viruses, considerably less effort has been invested in the area of non-icosahedral viruses. There are several notable examples of theoretical studies on non-icosahedral viruses. Keef *et al*. used viral tiling theory to describe the assembly of tubular viral particles with caps in the family of papovaviridae (11). A statistical mechanical model has also been developed to explain the *in vitro* equilibrium aggregation behavior of the coat proteins of the rigid rod–like tobacco mosaic virus(TMV) (12). The principles of a fullerene cone were also proposed to explain the construction of the HIV conical core (13). In addition, Nguyen *et al*. argued that relative stabilities of various virus capsid shapes such as spherocylindrical, spherical and conical shapes may be understood in terms of the continuum mechanics of closed hexagonal lattices (14).

The present work is inspired by the variety of precisely packed, elongated virus capsids, such as aberrant flock house virus (FHV), alfalfa mosaic virus (AMV) mutants, and bacteriophages $\phi$29 and T4. The principles of icosahedral virus capsid assembly have been addressed in our previous work (1). Our key findings were that the structures arising from the self-assembly of spheres into icosahedral viral capsids, as well as from the evaporation-driven assembly of colloidal spheres and the self-assembly of conical particles, result from free energy minimization subject to certain convexity constraints. For large spherical viruses, we showed that the precise packings arise with the initial help of a scaffold sphere that is subsequently removed, which agrees with the prediction made by continuum elasticity theory on viral capsid assembly (14). The convexity constraints are likely a natural outcome of the pressure of the encapsulated genomic materials and/or the scaffold proteins.

In the present work, we focus on the precise structures of prolate virus capsids and we argue that packings of hard, attractive spheres on a prolate spheroid surface can be used as a phenomenological model to explain the precise structures found in several elongated, prolate virus capsids. We believe this is the first simulation effort on the assembly of prolate virus capsid-like structures. We emphasize this work is not intended as an exhaustive study of optimum packings of attractive hard spheres on a prolate spheroid surface, nor is it intended as a realistic model of the actual virus assembly process. Instead, it should be considered as an extension of our previously developed general approach to provide insight into the geometrical packing problem as well as possible mechanisms for the assembly and structural origin of elongated, prolate virus capsids.

## METHODS

We perform Monte Carlo simulations of hard spheres interacting with an attractive square well potential and constrained to a prolate spheroidal surface. Applying this constraint involves moving the particles on the surface without bias, which requires us to generate random points uniformly on an ellipsoidal surface. To accomplish this, we modified the standard method of generating random points on a spherical surface. This method, and our modification to elliptical surfaces, is described below.

### Generating random points uniformly on a spherical surface.

The so-called "rejection" method and "trig" method are two of the most commonly used methods to generate random points uniformly on a spherical surface. The simplest to implement is the rejection method, which proceeds as follows (15). First, random points are generated inside a cube with an edge of length $l = 2$ by generating random values for the Cartesian coordinates x, y and z on [-1, 1]. Points that fall outside of the unit sphere inscribed within the cube are rejected, and those that fall inside are accepted. The vector of the accepted point is then scaled to the unit sphere surface. Because the sphere to cube volume ratio is $\pi/6$, the acceptance ratio of trial vectors is about 50%.

Although the rejection method is straightforward, the trig method, which may be used only in 3D, is faster because it does not reject any trial points. In this method, a unit sphere is represented in spherical coordinates as $x = \cos\phi\sin\theta$, $y = \sin\phi\sin\theta$ and $z = \cos\theta$, where $\phi$ is the azimuthal angle between 0 and $2\pi$, and $\theta$ is the polar angle between 0 and $\pi$. Although it is tempting to generate random values for $\phi$ and $\theta$ in order to calculate the position of a point on the surface of a sphere, this will create a nonuniform distribution of points on the surface because the three Cartesian coordinates are not independent. However, it suffices to use the following two random, independent variables, $\cos(\theta)$ (the z coordinate) and $\phi$, to generate uniformly distributed points on the spherical surface. The implementation of the trig method is therefore as follows (16):

1. Choose $u$ uniformly distributed on [-1,1].

2. Choose $\phi$ uniformly distributed on [0, $2\pi$).

3. Let $v = \sqrt{1-u^2}$.

4. Let $x = v\cos(\phi)$.

5. Let $y = v\sin(\phi)$.

6. Let $z = u$.

**Generating random points uniformly on an ellipsoidal surface.**

If we use the trig method as described above to generate points on a prolate spheroid surface by using the relationships $x = a\cos\phi\sin\theta$, $y = b\sin\phi\sin\theta$ and $z = c\cos\theta$ where $a$, $b$ and $c$ are three axes of an ellipsoid ($a = b$ for the prolate spheroid), the distribution of points will not be uniform and the number density of points on the surface will be higher at the ends of the long (major) axis than that at the ends of the short (minor) axis. We must therefore modify the trig method to eliminate this bias.

Williamson developed a general technique, upon which our approach is based, to generate uniformly distributed random points on a bounded smooth surface (17). The basic idea underlying this rejection-sampling method is to use a conventional method (such as the trig method) to generate random points on a spherical surface and then systematically reject some points in order to make the distribution uniform on a nonspherical surface, such as a prolate spheroid surface (after which it is then nonuninform on a spherical surface). Specifically, to generate a uniform distribution of points on a prolate spheroid surface, we (1) randomly generate a point on a spherical surface using the trig method, (2) calculate the trial position on the prolate spheroid surface, and (3) accept the point only when $g/g_{max} \geq \zeta^*$, where $g$ is the probability density function and $\zeta^* \in (0,1)$ is a uniformly distributed random number. The expression of the correction factor, $g/g_{max}$ is obtained as follows. The governing equation for an ellipsoid in Cartesian coordinates is $(\frac{x}{a})^2 + (\frac{y}{b})^2 + (\frac{z}{c})^2 = 1$. The ellipsoid is a sphere when $a = b = c$, and it is a prolate spheroid when $a = b < c$. For the prolate spheroid we consider here, we have $g/g_{max} = a\left(\frac{x^2+y^2}{a^4} + \frac{z^2}{c^4}\right)^{1/2}$ (see the Appendix for more details and a derivation). It is easy to verify that for $a = b = c = R$ (i.e., a sphere), $g/g_{max} = 1$, so all points will be accepted and the method reduces to the usual method of generating random points on a spherical surface. With this correction factor, the sampling is now biased in a way that fewer points will be accepted towards the ends of the long axis of a prolate spheroid, thereby generating a uniform distribution of points on a prolate spheroid surface.

In our simulations, the approach is implemented as follows:

1. Choose $u$ uniformly distributed on [-1,1].
2. Choose $\phi$ uniformly distributed on [0, $2\pi$).
3. Let $v = \sqrt{1-u^2}$.
4. Let $x = av \cos(\phi)$.
5. Let $y = bv \sin(\phi)$.
6. Let $z = cu$.
7. Accept the trial position $(x, y, z)$ if $g/g_{max} \geq \zeta^*$, where $g/g_{max} = a\left(\frac{x^2+y^2}{a^4} + \frac{z^2}{c^4}\right)^{1/2}$ and $\zeta^*$ is a uniformly distributed random number on [0, 1].

To check the validity of this method, we output the number density of points along the $z$ direction (also the direction of the long axis of a prolate spheroid). For $a = b = 4.5$, $c = 9.0$, we generate one million points randomly on a prolate spheroid surface using the rejection algorithm, i.e. the trig method with the correction factor $g/g_{max}$ as described above. We analyze the point density along the long axis ($c$ axis) by collecting points in multiple equal-distance bins of unit length along the $z$ direction and calculating the strip surface area along the spheroid long axis.

In Figure 1, red symbols represent the number densities of points along the $z$ axis from the trig method without using the correction factor $g/g_{max}$. Green symbols represent the number densities of points along the $z$ axis using the rejection algorithm. The blue line represents the expected overall number density of points, calculated by dividing the total number of generated points by the total surface area of a prolate spheroid. The relative standard deviation of the densities among 18 bins from our rejection method is only about 0.3%, showing excellent uniformity of the point distribution on the prolate spheroid surface.

The expression of $g/g_{max}$ used in this work is different from the one in the original reference (17). Although both approaches generate random points uniformly on a prolate spheroid surface, upon comparing the two expressions, we find our expressions to lead to an algorithm twice as fast as that by Williamson. The major difference is that our expression leads to a roughly 85% acceptance ratio of trial vectors, whereas the use of equation (17) in reference (17), which results in a much more complicated expression for $g/g_{max}$, has only a roughly 42% acceptance ratio.

**Simulation method.**

We perform Metropolis MC simulations with a convex-hull searching algorithm to study the assembly of hard, attractive spheres on a scaffold prolate spheroid. Each sphere represents a protein capsomer and the attractive interaction between spheres can be thought of as an effective attraction arising from averaging over the many hydrophobic sites on a generic capsomer. We first randomly distribute the hard spheres with square-well attraction (the interaction range, i.e., the width of the square well, $\lambda = 0.2\sigma$, where $\sigma$ is the particle diameter) on the surface of a large fictitious prolate spheroid surface with various aspect ratios (which can be obtained from experimental measurements of the virus capsid of interest). The surface is then gradually reduced in diameter to bring the particles together. At each iteration, a randomly chosen particle attempts to move radially or on the surface by a small amount. The ratio of radially inward to radially outward moves = 9 : 1, and the ratio of radial moves to surface moves = 2 : 8. The trial surface moves are generated using the previously described method of generating random points on the ellipsoid surface, which then are either accepted or rejected according to the standard Metropolis importance-sampling Boltzmann criterion if the trial moves do not violate either the excluded volume constraint between spheres or an imposed convexity constraint, described below.

The convexity constraint imposes a requirement on the convexity of the assembled structures that constrains all particles to the surface of the convex hull formed by the particles *at all times*, and is implemented with the "incremental algorithm" (18) from computational geometry. Any trial moves resulting in a structure violating the convexity criterion are rejected. The scaffold prolate spheroid provides all the particles on its surface an effective convex hull although the general convexity constraint is always invoked. After the scaffold prolate spheroid shrinks to a size smaller than the convex hull of the particles, it is the convexity constraint that further maintains the convex shape of the cluster in the simulation until the final structure appears. Simulations of attractive hard spheres on a shrinking prolate spheroid surface without a convexity constraint result in either concave or disordered structures that do not resemble any prolate virus structures if the scaffold is reduced to a point. Many independent runs are executed to ensure that the simulated structures are robust and can be repeatedly obtained.

**RESULTS**

We conjecture that the packing of hard, attractive spheres on a prolate spheroid surface is related to the precise structures of prolate virus capsids, just as the packing of hard, attractive spheres on a spherical surface can be related to the precise structures of icosahedral virus capsids (1, 4, 5). The objective of this work is to test this conjecture and reproduce the precise packings found in specific prolate virus capsids using simulation.

With the appropriately sized prolate spheroid as the scaffold, a series of precisely packed structures are obtained in our simulations and can be easily compared with several known prolate virus capsids. The simulated prolate structures are robust for a given size of the scaffold prolate spheroid. Since our particles represent capsomers rather than protein subunits, the correct number of capsomers is crucial to obtaining precisely packed structures. This information may be obtained from experimental measurements on prolate virus particles.

**Prolate structure with $N = 15$**

A hypothetical structure for the aberrant FHV (flock house virus) particle 1 (19) with an aspect ratio of 1.35 is shown in Figure 2(a). Our simulated structure obtained with a scaffold prolate spheroid of the same aspect ratio is shown in Figure 2(b), and we observe that the particle packing is identical to the packing of capsomers in the virus. This 15-capsomer capsid is the smallest possible prolate capsid. Note that the line of separation between the two halves of the capsid zigzags and that the caps are rotated relative to each other by 60°.

FHV is a $T = 3$ ($T$ = triangulation number; see below) icosahedral virus with 180 protein subunits arranged into 32 capsomers, which include 12 pentamers and 20 hexamers. Assembly of viral capsids is a highly precise process where erroneous or aberrant structures rarely occur. However, it was found that by deleting the molecular switch N-terminus whose positive charges are needed to neutralize the negatively charged RNA that is encapsidated into virions from the coat protein of FHV, multiple types of particles were observed which are smaller than the native $T = 3$ particle but larger than $T = 1$ particles (19). Some of the assembly products have short oval or ellipsoidal shape and resemble short bacilliform structures similar to those observed for the alfalfa mosaic virus (AMV). Besides the prolate capsid and icosahedral $T = 3$ capsids, they also found ill-defined capsid structures without precise packing.

Three kinds of aberrant FHV capsids are found in the experiments. All three have constant diameter of 20 nm and a length of 23, 27 and 30 nm, respectively. The size 20 × 23 corresponds to a $T = 1$ spherical capsid with icosahedral symmetry. The other two have elongated capsids with aspect ratios of 1.35 and 1.50.

It has been hypothesized that the assembled structures should have similar morphology or architecture to the AMV capsid, a prolate cylinder-like structure composed of multiple rings of hexamers capped at each end by one-half of a $T = 1$ icosahedral shell (19). Specifically, it was proposed that particle 1 with aspect ratio 1.35 has one extra ring containing three hexamers in addition to the $T = 1$ icosahedral structure, and total number of capsomers given by 12 + 3 = 15. Our simulated prolate structure containing 15 capsomers, which is in excellent agreement with the hypothetical structure, provides strong support for the proposed precise packing.

Dong *et al.* also proposed that aberrant FHV particle 2 with aspect ratio 1.50 contains two extra rings of hexamers (19) and the total number of capsomers can be calculated as 12 + 2×3 = 18. The comparison between their proposed structure and our simulation result at $N = 18$ will be discussed later.

**Prolate structure at $N = 17$ ($T = 1$, $Q = 2$)**

There are several conventions concerning the geometry of prolate structures. Some prolate particles can be characterized by two numbers: triangulation number T and elongation number $Q$ (in the literature, these are also referred to as $T_{end}$ and $T_{mid}$ (20)), just as icosahedral virus capsids can be characterized by the triangulation number T according to the "quasi-equivalence" theory in which each protein subunit is assumed to be in a similar (not identical and thus quasi-equivalent) environment (21). For example, the smallest icosahedron has 12 equilateral triangles on its surface and possesses 5-3-2 rotational symmetries through its 5-fold vertices, 3-fold face centers and 2-fold edges. For large icosahedral viruses, each of their 12 equilateral triangles can be considered as consisting of $T$ small triangles, where $T$ = 1, 3, 4, 7, 9, 12, 13, and so on. The total number of subunits is then $60T$. For elongated (prolate) capsids, the total number of protein subunits is instead calculated by the formula $N = 30(T+Q)$. An icosahedral particle has $Q = T$ and a prolate particle has $Q > T$.

Figure 3(a) illustrates the construction of an elongated $T = 1$, $Q = 2$ particle (22) and Figure 3(b) shows that our simulation yields the same $T = 1$, $Q = 2$ prolate structure containing 17 building blocks. The scaffold prolate spheroid used in the simulation has a size ratio of 1.30. Note that the two halves of the $T = 1$ icosahedral particle are in the eclipse position, i.e., the two caps are not in their native position as they are in the $T = 1$ icosahedral virus capsid. Instead, they are rotated relative to each other by 36 degrees and a row of 5 hexamers is inserted between the two hemispheres. The prolate structure has a 5-fold rotational symmetry.

**Prolate structure at $N = 18$**

Figure 4 shows (a) the hypothetical structure for the aberrant FHV capsid 2 (19) with an aspect ratio of 1.50 based on experimental studies and (b) our simulated prolate structure with a scaffold prolate spheroid of aspect ratio 1.60. We find identical structural characteristics between the simulated result and the hypothetical structure. Note that this 18-capsomer prolate structure can be visualized as having a zigzag row of 6 hexamers inserted between two halves of a $T = 1$ icosahedral structure, in contrast to a linear row of 3 and 5 structures inserted for the $N = 15$ and $N = 17$ prolate structures. Also, the two halves in the $N = 18$ prolate structure are in the correct position without rotation as compared to the corresponding $T = 1$ icosahedral structure. A similar structure was also proposed for one component of AMV (23) where the icosahedron is cut along the three-fold axis, giving the bacilliform particle three-fold rotational symmetry aligned along its long axis. The same 3-fold rotational symmetry is also observed for $N = 15$ and $N = 18$ prolate structures in our simulation.

**Prolate structure at $N = 42$ ($T = 3$, $Q = 5$ of bacteriophage ϕ29 capsid)**

Bacteriophages are viruses that infect bacteria instead of cells, and large phages usually have a tail to help with this process. While phage tails may differ in complexity, phage heads share many common features (20). A bacteriophage can extend its phage head to increase the volume by adding an extra tubular part to the center of an icosahedral capsid. The capsids of bacteriophages can take diverse shapes like filaments (phage M13), spherical icosahedral or isomeric shape (phage P1), and prolate shape (phage ϕ29 and T4). Prolate bacteriophages ϕ29 and T4 have been investigated the most. In pioneering work by Tao *et al.*, the assembly, genome release and surface structure of bacteriophage ϕ29 have been revealed (24). Both the prohead and matured phage head of bacteriophage ϕ29 have a prolate shape. The surface packing of bacteriophage ϕ29 capsid has been shown as a $T = 3$, $Q = 5$ prolate structure (24) and the construction principles of this elongated virus capsid have also been discussed (22). However, no simulation study of the assembly process has been reported for this virus.

The prolate head of bacteriophage ϕ29 as revealed by experiments contains 42 capsomers and has a dimension of 42 nm × 54 nm and thus an aspect ratio of 1.286. Figure 5(a) and (b) show the proposed surface packing of the bacteriophage ϕ29 capsid and Figure 5(c) shows our simulated structure at $N = 42$ using a scaffold prolate spheroid of aspect ratio 1.34. Again, we find perfect agreement between the simulation and the experimental observation. The elongated structure is most easily visualized as dividing a $T = 3$ icosahedral structure into two hemispheres, and then inserting a row of ten hexamers between the two hemispheres. Note the zigzag pattern again as found in the case of the $N = 18$ prolate structure.

**DISCUSSION AND CONCLUSIONS**

We demonstrated that the assembly of hard spheres with square-well attraction on a prolate spheroid surface leads to identical packings as those found in several elongated, prolate viruses. Here we discuss some implications of our simulation results.

This work can be considered as an example of the more general problem of confined assembly on non-spherical surfaces. By comparison with our previous work (1), our simulations show that confinement on a prolate spheroid surface is directly responsible for the precise packing of the resulting structures and the symmetries that appear in elongated viral capsids. The sizes of virus particles investigated in this work are small due to the computational demands of self-assembling larger structures, such as the bacteriophage T4 capsid, which contains about 172 capsomers. The structure of bacteriophage T4 has not been resolved unambiguously. Some conflicting arguments have been raised in different experimental works. For example, two candidate structures, $T = 13, Q = 20$ (25) and $T = 13, Q = 21$ (26) have been proposed for this virus, though the former is now more favored. Future simulations may provide additional insight into the likely structure. In addition, TMV can be considered as a very long prolate (rod-shaped) virus, and thus might be simulated by our approach. As such, improving the efficiency of our current approach so as to extend it to the problem of large prolate viruses and rod-like viruses is planned.

One may speculate as to the physical origin of the prolate ellipsoidal constraint used in the simulation, which we found is necessary to guarantee the successful assembly of hard spheres into the correct structure. A possible source of this constraint is the presence of scaffold proteins. Both phage φ29 and T4 need scaffold proteins to assemble coat proteins into capsids correctly (24, 25). The capsid assembly process can be nucleated by scaffold proteins. For large phages, an elongated "core" is essential for assembly of the prolate T-even head, but usually an inner scaffold (core) without precise structure suffices because such a disordered scaffold structure may adequately define the shell radius or volume, the kind of information that the assembling protein subunits may utilize to determine their final precise configuration in capsids (20). Genomic DNA/RNA can also play a similar role as scaffold proteins, as found in bacteriophage T4. In AMV particles, the different lengths of the RNA molecules dictate the lengths of the capsid. As such, a similar assembly motif as that in the assembly of spherical icosahedral viruses may be at work (27). In that motif, scaffold proteins or nucleic acids induce the capsomers to assemble into a prolate ellipsoid-like structure around the genomic material as a first step in capsid formation. As the nucleic acid is neutralized by the amino terminal tails on the assembling proteins, the genomic material shrinks, pulling the capsid proteins in tighter until they eventually restructure into their ultimate symmetry. This may warrant our use of a scaffold prolate spheroid in assembling hard spheres on the prolate spheroid surface to reach the expected precise structures, and thus may provide a phenomenological viewpoint about how prolate virus capsids assemble.

In the simulations, we find that if we reduce the scaffold prolate spheroid to a point without enforcing a general convexity constraint, we can only obtain either concave or disordered structures that do not resemble any prolate virus structures. If we reduce the scaffold prolate spheroid to a point while maintaining the general convexity constraint, prolate structures at $N = 17$, 18 and 42 can be maintained but no prolate structure can be obtained at $N = 15$. Interestingly, for small prolate structures at $N = 15$, 17 and 18, the use of a scaffold prolate spheroid is sufficient for the appearance of the desired precise packings as long as we stop reducing the scaffold's size at certain point. However, for prolate structure at $N = 42$, a convexity constraint is necessary to obtain the final packing and the polyhedral shape of the structure after the scaffold shrinks to a size that is smaller than the structure itself. Thus the combination of the prolate surface and convexity is crucial for the self-assembly of attractivehard spheres into elongated capsid-like structures. This knowledge should prove useful in understanding the assembly of prolate virus capsid structures as well as the self-assembly of synthetic particles and the design of artificial viruses, nanoshells based on bacteriophage capsids (28), and the development of antiviral drugs.

## APPENDIX

According to the general sampling procedure using rejection algorithm by Williamson (17), the correction factor, $\frac{g}{g_{max}} = \frac{|n(u,v)|}{|n(u,v)|_{max}} = \frac{|\nabla f|}{|\nabla f|_{max}}$ for surfaces governed by $\{(x,y,z) | f(x,y,z) = c\}$, where $n(u,v)$ is the normal vector to the surface that is to be sampled, $c$ is a constant and $\nabla f$ is the gradient. For the prolate spheroid considered in this work, we have

$$|\nabla f| = (\frac{x^2 + y^2}{a^4} + \frac{z^2}{c^4})^{1/2} \qquad (1)$$

Because $x^2 + y^2 = a^2(1 - z^2/c^2)$, we substitute this equation into the above equation and rearrange, yielding,

$$|\nabla f| = [\frac{a^2(1-z^2/c^2)}{a^4} + \frac{z^2}{c^4}]^{1/2} = [\frac{1}{a^2} + \frac{z^2}{c^2}(\frac{1}{c^2} - \frac{1}{a^2})]^{1/2} \qquad (2)$$

Because $\frac{1}{c^2} - \frac{1}{a^2} < 0$, $|\nabla f|$ reaches a maximum at z = 0, i.e. $|\nabla f|_{max} = \frac{1}{a}$. Finally, we have

$$\frac{g}{g_{max}} = \frac{|\nabla f|}{|\nabla f|_{max}} = a(\frac{x^2 + y^2}{a^4} + \frac{z^2}{c^4})^{1/2} \qquad (3)$$

It is obvious that when $a = c$ (the limiting case of sphere), $\frac{g}{g_{max}} = 1$ and then all trial moves will be accepted, and the approach reduces to the original trig method that is used to generate random points on a spherical surface.

## ACKNOWLEDGEMENT

This work was supported by the US Department of Energy, Grant No. DE-FG02-02ER46000.

**FIGURE CAPTIONS**

**FIGURE 1** Number density of points obtained by method of generating random points on a prolate spheroid surface with (solid green circles) and without (solid red circles) the correction factor $g/g_{max}$, and the calculated average number density (blue line).

**FIGURE 2** $N = 15$, the smallest prolate structure. (a) Hypothetical structure for aberrant FHV (flock house virus) particle 1 with an aspect ratio of 1.35. Reprinted from reference (19). (b) Simulated prolate structure with scaffold prolate spheroid of aspect ratio 1.35. In virus capsids, a hexamer is a capsomer surrounded in its nearest neighbor shell by six other capsomers and consists of six protein subunits. A pentamer is a capsomer surrounded by five other capsomers and consists of five protein subunits. The assembly contains one extra ring of three spheres representing hexamers (in gray) and 12 spheres representing pentamers shown in gold.

**FIGURE 3** $N = 17$, $T = 1$, $Q = 2$ prolate structure. (a) Physical model for the principles of elongated capsid construction for $T = 1$, $Q = 2$. Reprinted from reference (22). (b) Simulated prolate structure with scaffold prolate spheroid of aspect ratio of 1.30. The assembly contains one linear ring of 5 spheres representing hexamers (in gray) and 12 spheres representing pentamers shown in gold.

**FIGURE 4** $N = 18$ prolate structure. (a) Hypothetical structure for aberrant FHV (flock house virus) particle 2 with aspect ratio 1.50. Reprinted from reference (19). (b) Simulated prolate structure with scaffold prolate spheroid of aspect ratio 1.60. Two halves of $T = 1$ icosahedral cap with two rings of spheres representing hexamers (each ring has 3 hexamers), or one belt of six spheres representing hexamers in zigzag pattern (in gray). 12 spheres representing pentamers are shown in gold.

**FIGURE 5** $N = 42$, $T = 3$, $Q = 5$ prolate structure. (a) (left) Triangulation net of $T = 3$, $Q = 5$ for the (right) prolate prohead of bacteriophage $\phi 29$. Reprinted from reference (24). (b) Physical models for (left) an icosahedral capsid with $T = 3$ quasi-symmetry; (middle) a prolate $T = 3$, $Q = 5$ capsid; (right) visualization of the elongated capsid with two icosahedral caps and the insertion of a row of ten hexamers in a zigzag pattern. Reprinted from reference (22). (c) Simulated $T = 3$, $Q = 5$ prolate structure with scaffold prolate spheroid of aspect ratio 1.34. Two halves of $T = 3$ icosahedral cap with a row of 10 spheres representing hexamers in zigzag pattern (in gray). 12 spheres representing pentamers are shown in gold.

**FIGURE 1**

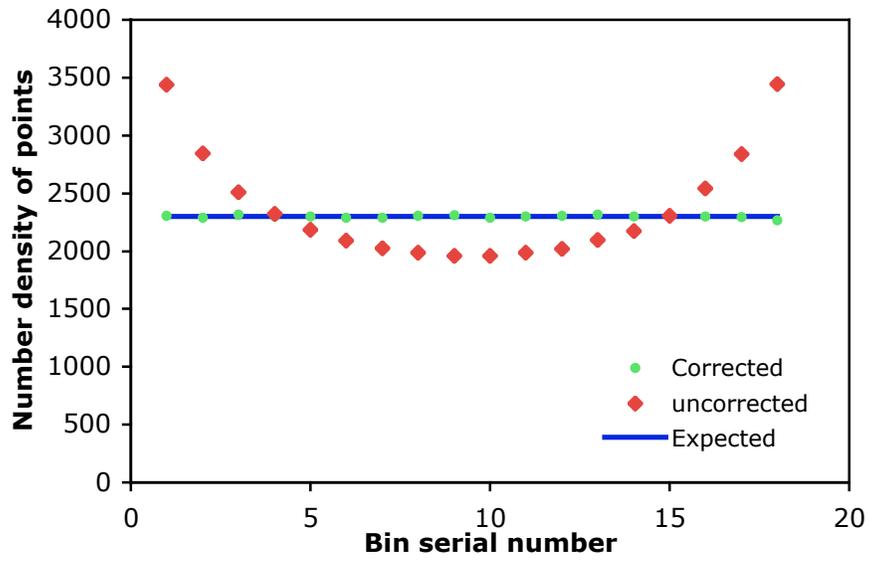

**FIGURE 2**

(a)

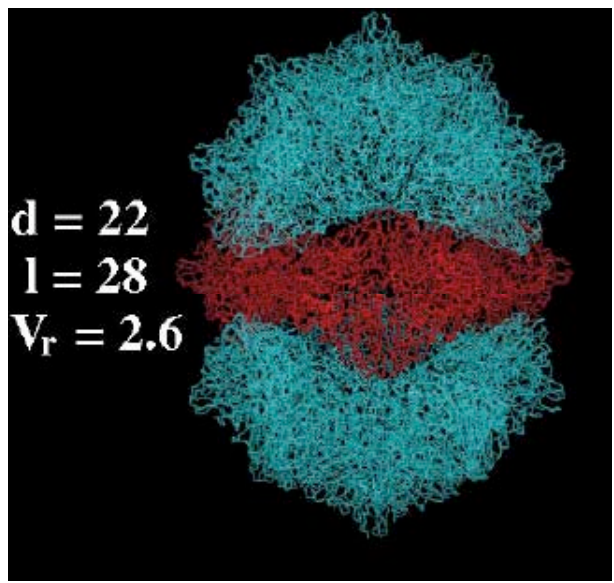

(b)

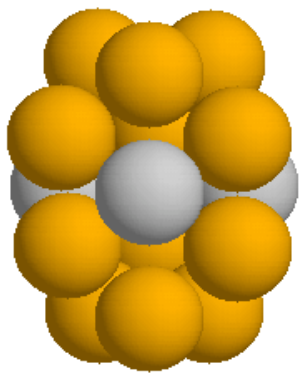

**FIGURE 3**

(a)

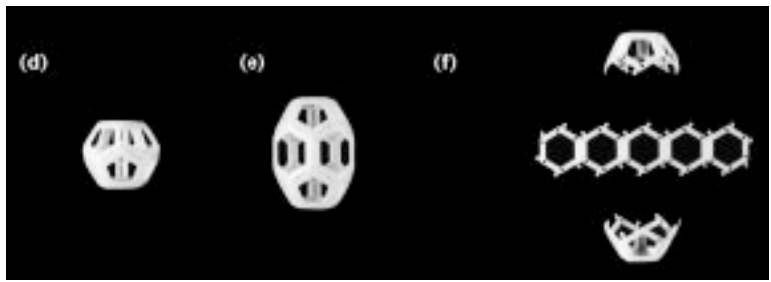

(b)

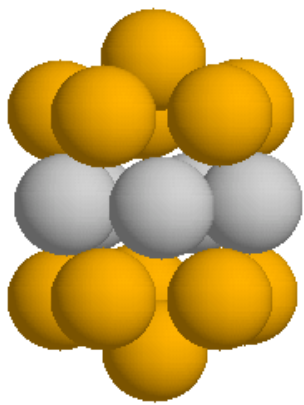

**FIGURE 4**

(a)

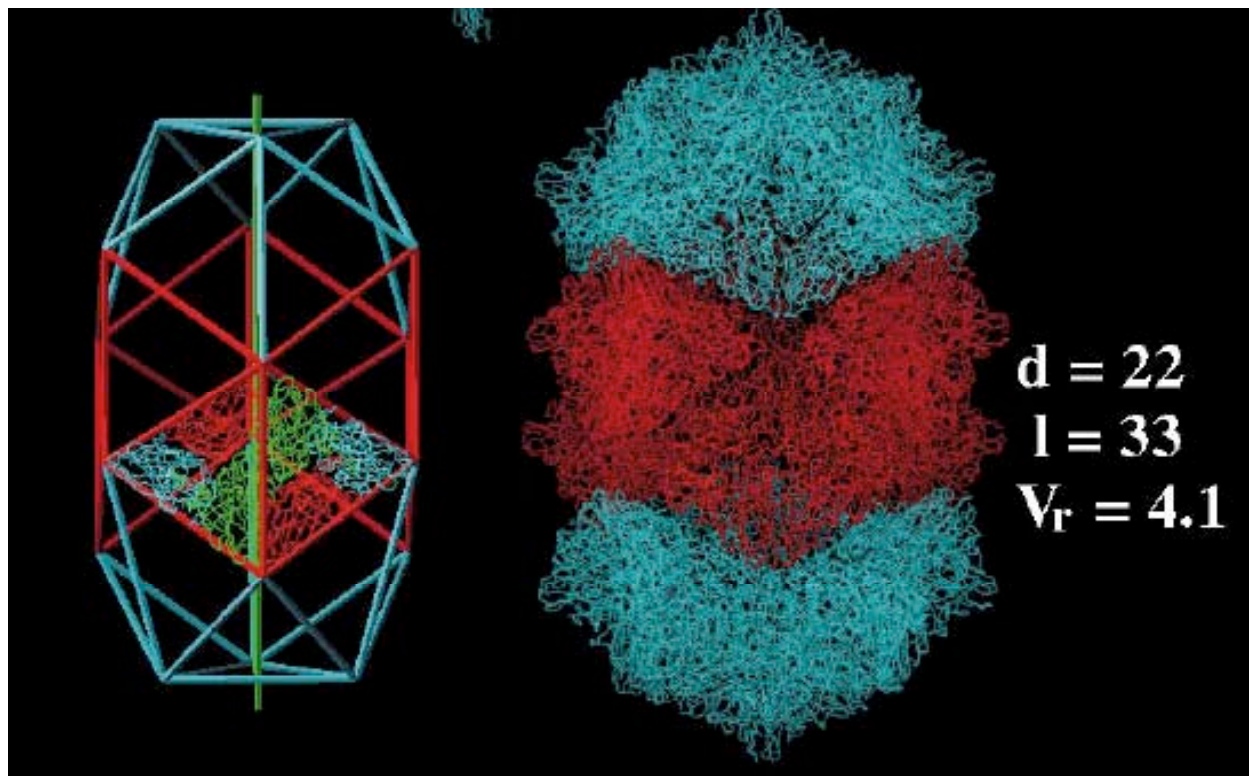

(b)

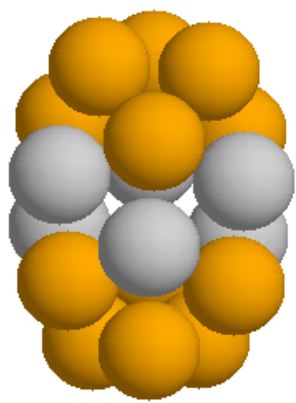

**FIGURE 5**

(a)

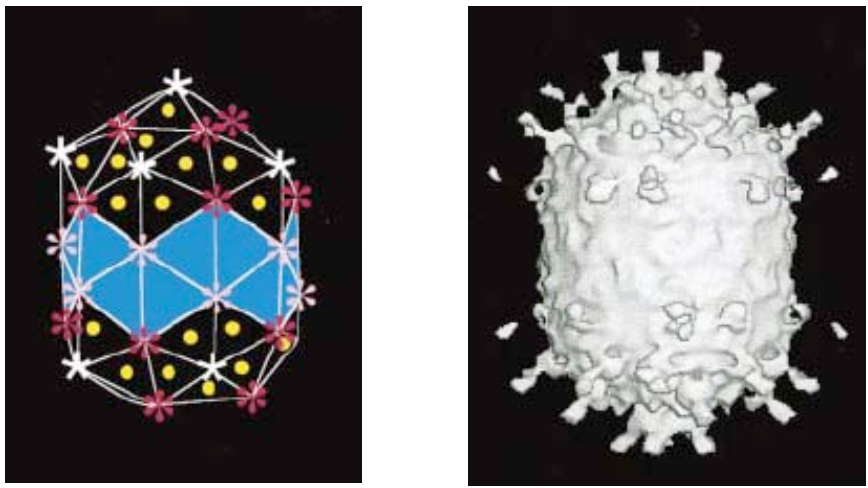

(b)

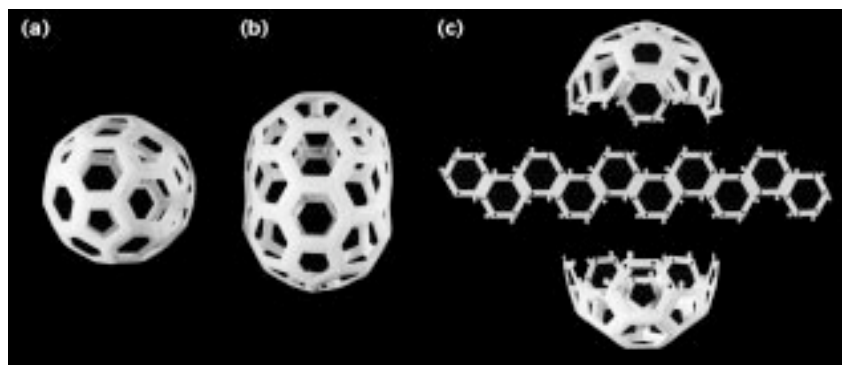

(c)

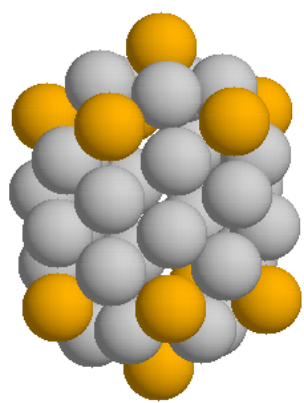